# Ferromagnetism in Mn Doped GaN:

# From Clusters to Crystals

G. P. Das,[*] B.K. Rao, and P. Jena

Physics Department, Virginia Commonwealth University

Richmond, VA 23284-2000

## Abstract

The magnetic coupling between doped Mn atoms in clusters as well as crystals of GaN has been studied from first principles using molecular orbital theory for clusters and linearized muffin tin orbitals-tight binding formulation (LMTO-TB) for crystals. The calculations, based on density functional theory and generalized gradient approximation for exchange and correlation, reveal the coupling to be ferromagnetic with a magnetic moment ranging from 2.0 to 4.0 $\mu_B$ per Mn atom depending on its environment. Mn atoms also tend to cluster and bind more strongly to N atoms than to Ga atoms. The significant binding of Mn to GaN clusters further indicates that it may be possible to increase the Mn concentration in GaN by using a porous substrate that offers substantial interior surface sites.

[*] On leave from Bhabha Atomic Research Centre, Mumbai, India

The discovery of ferromagnetism in Mn doped InAs and GaAs with a Curie point of 110K[1] and the subsequent theoretical prediction[2] that the Curie point in Mn doped GaN could be higher than the room temperature have created an intense interest in the study of dilute magnetic semiconductors (DMS). Studies of these systems are driven not only by the academic interest in understanding the origin of ferromagnetism from a fundamental point of view but also by the fact that new semi-conducting devices that combine electron's charge and spin could be of high technological interest.

There are two central questions that need to be addressed in the quest for doped magnetic semiconductors with a Curie temperature above 300K: (i) What is the origin of the ferromagnetic coupling in these systems? (ii) How does one increase Mn concentration so that the magnetic ion density and consequently the Curie temperature ($T_c$) could be enhanced? Several attempts have been made in the recent years both experimentally and theoretically to address these issues. Overberg *et al.*[3] reported a $T_c$ between 10 and 25K in GaN samples containing 7% Mn while Theodoropolou *et al.*[4] and Reed *et al.*[5] have reported ferromagnetism in (GaMn)N with $T_c$ of 250 and 228-370K respectively. Sonoda *et al.*[6] succeeded in incorporating upto 9% Mn in GaN and estimated (by extrapolation of the magnetization vs. temperature curve using mean field approximation) a $T_c$ as high as 945K. Although the growth mechanism seems to play a vital role, the reason for such a wide variation of $T_c$ is not understood.

There have been several theoretical attempts based on model calculations to study this problem as well. The original explanation of ferromagnetism in DMS systems was given by Dietl *et al.*[2] in terms of hole-mediated Rudermann-Kittel-Kasuya-Yosida (RKKY) interaction. This approach, which implies that a Fermi surface must exist, has recently been questioned by Litvinov and Dugaev.[7] These authors, instead, propose that ferromagnetism in DMS systems is due to localized spins in the magnetic impurity acceptor level of the semiconductor crystal, that excite band electrons due to s-p or p-d exchange interaction. Few first principles calculations that examine if binding of Mn is energetically favorable, how their magnetic moments are coupled, and if this coupling depends on the environment of the Mn atoms are available.

In this paper we present the results of first principles theoretical calculations of the electronic structure, energetics, and magnetism of Mn doped GaN in various structural forms that simulate binding of Mn on to surface as well as bulk sites. We investigate if Mn substitution is energetically favorable and if its binding energy depends on the environment. We also determine the charge and spin state of Mn and the coupling between the spins at Mn sites. We have done this by doping a pair of Mn atoms into $(GaN)_x$ ($x \leq 3$) clusters (surface sites) as well as into wurtzite GaN crystal (bulk sites). This allows us to study if the coupling between Mn atoms is ferrmagnetic or antiferromagnetic. Since the environment around Mn sites changes significantly with cluster size as well as in the crystal, we are able to determine the effect of local bonding on the energetics, electronic structure, and magnetic properties of doped Mn. We find that the Mn atoms are coupled ferromagnetically irrespective of the hosts we have considered. This is particularly interesting since bulk Mn is antiferromagnetic while in very small clusters the coupling is ferro- and/or ferrimagnetic.

The calculations on clusters were carried out by using the linear combination of atomic orbitals - molecular orbital (LCAO-MO) method. The atomic orbitals centered at individual Ga, N, and Mn sites were represented by gaussian orbitals. We used the 6-311 G** basis set available in the Gaussian 98 code.[8] The total energies were calculated using the density functional theory (DFT) and generalized gradient approximation (GGA) for exchange and correlation.[9] The geometries were optimized by calculating the force at every atomic site and relaxing the geometry until the forces vanish. The threshold for this was set at 0.000450 a.u./Bohr. Since Mn atom could carry a magnetic moment, the geometries were optimized for various spin multiplicities, $M = 2S+1$ to arrive at the ground state. We first discuss the structure and properties of $(GaN)_x$ clusters and study the manner in which they are altered by Mn doping.

In Fig. 1 we present the geometries of $(GaN)_x$ clusters on the left column and those of $(GaN)_xMn_2$ clusters on the right column. Some of the representative bond distances are marked in the figure. The Ga-N distance in the dimer is 1.88Å and changes only slightly (~0.1Å) as clusters grow. Note that the nearest distance between Ga and N in the wurtzite crystalline GaN is 1.95Å. These close values between inter-atomic distances in clusters and crystals is characteristic of covalently bonded systems. As the $(GaN)_x$ clusters are doped with Mn atoms, the structures change significantly. The GaN bond distances get enlarged by almost 1Å in going from GaN to $(GaN)Mn_2$, but this enhancement decreases rapidly in larger $(GaN)_xMn_2$ clusters yielding a value of about 2.4Å in $(GaN)_2Mn_2$ and 2.05Å in $(GaN)_3Mn_2$. Since this GaN bond distance is close to that in GaN crystal, namely 1.95Å, it indicates that doping of Mn into clusters may illustrate the salient features of the electronic structure of bulk Mn doped GaN. We also note from Fig. 1 that Mn-Mn distance in $(GaN)_xMn_2$ clusters vary from 3.11Å in $(GaN)Mn_2$ to 2.65Å in $(GaN)_3Mn_2$. In bulk α-Mn, the Mn-Mn distances also vary over a wide range, namely between 2.25Å and 2.95Å.

We now discuss the energetics of these clusters. The binding energy of $(GaN)_x$ clusters is defined as

$$E_b = [xE(GaN) - E(GaN)_x]/x \qquad (1)$$

We define the energy gain in adding a GaN dimer to an existing $(GaN)_{x-1}$ cluster as

$$\Delta E_0 = E(GaN) + E[(GaN)_{x-1}] - E[(GaN)_x] \qquad (2)$$

Similarly, the energy gain in adding two Mn atoms to an existing $(GaN)_x$ cluster is defined as

$$\Delta E = E[(GaN)_x] + 2E(Mn) - E[(GaN)_xMn_2] \qquad (3)$$

Here E represents the total energy of the corresponding systems. The results are given in Table 1. We first note that the binding energy of GaN dimer measured against dissociation into Ga and N atoms is 2.18 eV. As associative GaN units are added, the binding energy in Eq.(1) steadily increases. On the other hand, the energy gain in adding successive GaN units (see Eq.(2)) first increases and then decreases indicating that $(GaN)_2$ is a relatively more stable unit.

Doping of Mn atoms to $(GaN)_x$ clusters is found to be energetically quite favorable. For example, addition of two Mn atoms to a GaN dimer results in an energy gain of 5.39 eV. It should be mentioned that the binding energy of an $Mn_2$ dimer is less than 0.1 eV as the Mn atom has a half filled 3d and filled 4s shell and hence interacts weakly with another Mn atom. The nature of bonding changes in the presence of GaN. Mn and N atoms form a strong bond due to charge transfer from Mn to N. As a matter of fact, the binding energy of MnN dimer is 3.07 eV which is significantly larger than that of GaN, namely 2.18 eV. In addition, the two Mn atoms that interact weakly with each other in $Mn_2$ due to their closed 4s shells, no longer do so in the presence of N. Their coupling is mediated by N. The fact that the bonding of MnN is stronger than that of GaN suggests that when Mn is deposited on GaN substrate, Mn can replace Ga atoms and cluster around N. This is confirmed by recent experiment of Prokes and coworkers.[10] Since small $(GaN)_x$ clusters represent all surface atoms, our results suggest that doping of Mn in GaN surfaces as well as porous GaN that contain large internal surfaces and voids is energetically favorable. The successive energy gains in adding two Mn atoms to $(GaN)_x$ clusters are also substantial although they tend to oscillate with cluster size.

We now consider the magnetic properties of these clusters. We have studied the energetics of these clusters by varying their spin multiplicities, $M = 2S + 1$. In Fig. 1 we list the total magnetic moments of the clusters for which the energy is the minimum. The details of higher energy isomers carrying different moments will be published elsewhere. The magnetic moments of free Ga, N, and Mn atoms are respectively $1\mu_B$, $3\mu_B$, and $5\mu_B$. The magnetic moments of clusters of $(GaN)_x$ are $2\mu_B$ for x = 1, 2, and $0\mu_B$ for x = 3. For those clusters that have finite magnetic moment, much of it is located at the N-site which is antiferromagnetically coupled to the moment at Ga site. As clusters increase in size, it is expected that the individual moments at Ga and N sites will decrease and eventually vanish since bulk GaN is non-magnetic. We already see this happen in clusters as small as $(GaN)_3$.

As the Mn atoms are doped, the $(GaN)_xMn_2$ (x ≤ 3) clusters exhibit substantial magnetic moments. For example, the total magnetic moments of $(GaN)Mn_2$ and $(GaN)_3Mn_2$ are $8\mu_B$ each. Most of these moments are localized at the Mn sites (4.10 $\mu_B$ in GaN and 3.39 $\mu_B$ in $(GaN)_3$), and the two Mn atoms are coupled ferromagnetically. The moments at Ga and N sites are very small and couple antiferromagnetically to those at Mn sites. In all these clusters the Mn-Mn distance is larger than 2.5Å. It has been known from studies of free Mn[11] and MnO[12] clusters that the coupling between Mn atoms could be antiferromagnetic if their interatomic distances are reduced. Thus, it is important that for Mn atoms to couple ferromagnetically, they need to be kept apart by more than 2.5Å. In bulk GaN this is not a problem as Mn substitutes the Ga site and the nearest neighbor distance between two Ga atoms in bulk GaN is 3.19Å. We will show in the following through LMTO-TB band structure calculations that this is indeed the case.

In order to study the effect of the Mn impurity on the electronic structure of GaN crystal and the interaction between Mn magnetic moments, we have considered the hexagonal wurtzite structure which lies lower in energy than the cubic zinc blende structure. A super cell which is eight times larger than the wurtzite GaN unit cell was constructed that accommodates 16 Ga and 16 N atoms. Two of the nearest Ga atoms were selectively replaced by Mn atoms so that the super cell formula unit becomes $Ga_{14}Mn_2N_{16}$. It should be noted, however, that this 32 atom super cell is one of the smallest super cell that ensures separation between the impurities in neighboring super cells by at least a few times the Ga-N bond length. Similar super cells have been used[13] for Be impurity in wurtzite GaN.

All the band structure calculations reported in this work have been performed using self-consistent tight-binding linear muffin-tin orbital (TB-LMTO) method with the Atomic Sphere Approximation (ASA) and the "Combined Correction".[14] We have used the local spin-density approximation to DFT, along with the gradient correction as per the original Perdew-Wang formulation.[15] The super cell is divided into space-filling and therefore slightly overlapping spheres centered on each atom. Since the wurtzite GaN is an open structure, we had to introduce two different types of empty spheres (2 of each type) in the unit cell of GaN, thereby making the total number of spheres in the wurtzite unit cell as 8. This translates to a 64 atom supercell with 32 real atoms and 32 empty spheres. All the calculations have been performed with the experimental lattice parameters (a = 3.189Å and c = 5.185Å), and no lattice relaxation effects have been taken into account. Using the so-called Hartree potential plot prescription, we have fixed the Ga and N atomic sphere radii to be 1.227Å and 1.015Å, which are roughly proportional to the corresponding covalent radii of 1.62Å and 1.26Å of Ga and N respectively. For Mn atomic spheres, we have used the same atomic sphere radius as that of Ga. Brillouin Zone (BZ) integration has been performed using the improved tetrahedron method.[16] In all our super cell calculations, we have used (6,6,4) **k**-mesh which corresponds to 84 **k**-points in the irreducible wedge of the simple cubic BZ. Spin-polarized scalar relativistic (i.e. without spin-orbit interaction which is not significant for GaN) calculations have been performed with minimal basis set consisting of s-, p-, and d-orbitals ($\ell = 2$) for Ga, Mn and N, with N-d orbitals downfolded. Note that the localized semicore 3d states of Ga have been treated as fully relaxed band states, as emphasized by other workers[13,17] also. Apart from the valence states of Ga, Mn and N, the core orbitals were kept frozen to their isolated atomic form.

We have first benchmarked our calculations by comparing the electronic, cohesive and structural properties of bulk Wurtzite GaN with those reported in the literature. The band structure shows a direct gap of ≈2 eV at the Γ-point, which can be artificially 'opened up' to match with the experimental gap, by applying some external $\ell$-dependent potential (as done by Christensen and Gorczyca[17] for calculating deformation potential and the optical properties etc.) The zero of the energy is fixed at the top of the valence band, which consists of s- and p-orbitals of Ga and N. The semi-core like 3d states appear as narrow bands ~10 eV below the Fermi level. The valence band widths Wl, W2, and W3 are found to be 7.2 eV, 2.6 eV and 0.7 eV respectively, which are in very good agreement with the orthogonalized LCAO results[18] on Wurtzite GaN. Both the overestimation of binding energy and the underestimation of band gap are typical of LDA that are partially salvaged by incorporation of GGA. In all our calculations, therefore, we have used GGA as discussed above. More rigorous (and also computationally demanding) GW calculations have been reported[19] that show improved energy gaps, due to incorporation of nonlocal exchange-correlation potential.

On introduction of a pair of Mn atoms in Ga-substitutional positions (Mn-Mn distance = 3.189Å) we observe several drastic changes in the electronic structure. The total and partial (Mn-projected) densities of states of $Ga_{14}Mn_2N_{16}$ (Fig. 2) show that the Fermi level passes through Mn d-bands for the majority spin. The minority spin Mn d-band lies above the Fermi level and merges with the bottom of the conduction band. The Mn d-band width is ~2.4eV. The peak of the majority spin Mn d-band lies ~1.8 eV above the top of the valence band of GaN. This conforms to the conventional wisdom of Mn acting as an effective mass acceptor ($d^5$ + h) and also with the recent deep-level optical spectroscopy measurements on lightly Mn-doped samples,[20] which indicates that Mn forms a deep acceptor level at 1.4eV above the GaN band gap. The $e_g$ and $t_{2g}$ peaks of Mn are both strongly spin-split. The coupling between Mn atoms is ferromagnetic and a localized magnetic moment of ~$3.5\mu_B$ is manifested on the Mn atoms. Some weak polarization is also observed at the nearest host atoms surrounding the impurity. This is in agreement with the results reported by Fong *et al.*[21] and by Schilfgaarde and Mrysov[19] from their LDA supercell calculations on zinc blende GaN doped with Mn. The later investigation also hinted at the formation of Mn-clusters in GaN, in which the ferromagnetic coupling strength decreases with increasing Mn-Mn separation (~3-5 Å). When Mn concentration is increased beyond a critical limit, the magnetic moment reduces drastically. Also we find that there is no ferromagnetic coupling if instead of Ga substitutional position, the Mn impurity goes to an interstitial position in the wurtzite lattice.

The above results indicate that the coupling between Mn atoms is ferromagnetic whether they are doped into the crystal or clusters. Equally interesting is our finding that the Mn atoms retain a magnetic moment of about $3.5\mu_B$ irrespective of their environment. Since clusters represent an extreme case of surface states and crystal sites represent substitutional bulk environment, we are convinced that doping of Mn in GaN whether they are porous, crystalline, or thin layers would lead to ferromagnetic coupling between Mn atoms. Our results further suggest that clustering of Mn around N is energetically favorable. The sensitivity of the measured $T_c$'s to experimental growth conditions may very well be due to the clustering[22] of Mn around N.

This work was supported by a grant from the Office of Naval Research under a Defense University Research Initiative on Nanotechnology (DURINT) and by a grant from the Department of Energy (DE-FG02-96ER45579).

## References

1. H. Ohno, Science **281**, 951 (1998)

2. T. Dietl, H. Ohno, F. Matsukura, J. Cibert, and D. Ferrand, Science **287**, 1019 (2000); T. Dietl, H. Ohno, and F. Matsukura, Phys. Rev. B 63, 195205 (2001); T. Dietl, F. Matsukura, and H. Ohno, cond-mat/0109245, 13$^{th}$ Sept, 2001.

3. M. E. Overberg, C. R. Abernathy, S. J. Pearton, N. A. Theodoropoulou, K. T. McCarthy, and A. F. Hebard, Appl. Phys. Lett. **79**, 1312 (2001).

4. N. Theodoropoulou, A. F. Hebard, M. E. Overberg, C. R. Abernathy, S. J. Pearton, S. N. G. Chu, and R. G. Wilson, Appl. Phys. Lett. **78**, 3475 (2001).

5. M. L. Reed, N. A. El-Masry, H. H. Stadelmaier, M. K. Ritums, M. J. Reed, C. A. Parker, and S. M. Bedair, Appl. Phys. Lett. **79**, 3473 (2001).

6. S. Sonada, S. Shimizu, T. Sasaki, Y. Yamamoto, and H. Hori, Cond-mat/0108159, 9$^{th}$ Aug 2001.

7. V. I. Litinov and V. K. Dugaev, Phys. Rev. Lett. **86**, 5593 (2001).

8. GAUSSIAN 98, Revision A.7, M. J. Frisch, G. W. Trucks, H. B. Schlegel, G. E. Scuseria, M. A. Robb, J. R. Cheeseman, V. G. Zakrzewski, J. A. Montgomery, Jr., R. E. Stratmann, J. C. Burant, S. Dapprich, J. M. Millam, A. D. Daniels, K. N. Kudin, M. C. Strain, O. Farkas, J. Tomasi, V. Barone, M. Cossi, R. Cammi, B. Mennucci, C. Pomelli, C. Adamo, S. Clifford, J. Ochterski, G. A. Petersson, P. Y. Ayala, Q. Cui, K. Morokuma, D. K. Malick, A. D. Rabuck, K. Raghavachari, J. B. Foresman, J. Cioslowski, J. V. Ortiz, A. G. Baboul, B. B. Stefanov, G. Liu, A. Liashenko, P. Piskorz, I. Komaromi, R. Gomperts, R. L. Martin, D. J. Fox, T. Keith, M. A. Al-Laham, C. Y. Peng, A. Nanayakkara, C. Gonzalez, M. Challacombe, P. M. W. Gill, B. Johnson, W. Chen, M. W. Wong, J. L. Andres, C.

Gonzalez, M. Head-Gordon, E. S. Replogle, and J. A. Pople, Gaussian, Inc., Pittsburgh PA, 1998.

9. A. D. Becke, Phys. Rev. A **38**, 3098 (1988); J. P. Perdew, K. Burke, and Y. Wang, Phys. Rev. B **54**, 16533 (1996); K. Burke, J. P. Perdew, and Y. Wang in *Electronic Density Functional Theory: Recent Progress and New Directions*, Ed. J. F. Dobson, G. Vignale, and M. P. Das (Plenum, 1998).

10. S. Prokes (Private communication).

11. S. K. Nayak, M. Nooijen, and P. Jena, J. Phys. Chem. A **103**, 9853 (1999); S. N. Khanna, B. K. Rao, P. Jena, and M. Knickelbein (to be published),

12. S. K. Nayak and P. Jena, J. Am. Chem. Soc. **121**, 644 (1999).

13. C. G. van de Walle, S. Limpijumnong, and J. Neugebauer, Phys. Rev. B **63**, 245205 (2001).

14. O. K. Andersen, Phys. Rev. B **12**, 3060 (1975); O. K. Andersen and O. Jepsen, Phys. Rev. Lett. **53**, 2571 (1984). We have used the latest version of the Stuttgart TB-LMTO-ASA program.

15. J. P. Perdew and Y. Wang, Phys. Rev. B **33**, 8800 (1986); J. P. Perdew, Phys. Rev. B **33**, 8822 (1986); J. P. Perdew and Y. Wang, Phys. Rev. B **45**, 13244 (1992).

16. P. Bloechl, O. Jepsen, and O. K. Andersen, Phys. Rev. B **49**, 16223 (1994).

17. N. E. Christensen and I. Gorczyca, Phys. Rev. B **50**, 4397 (1994)

18. Y. -N. Xu and W. Y. Ching, Physr. Rev. B **48**, 4335 (1993).

19. M. van Schilfgaarde and O. Mryasov, in Proc. of APS Meeting, Minneapolis, MN 2000, abstract V26; Phys. Rev. **B 63**, 233205 (2001).

20. R. Y. Korotkov, J. M. Gregie, and B. W. Wessels, Appl. Phys. Lett. **80**, 1731 (2002).

21. C. Y. Fong, V. A. Gubanov, and C. Boekema, J. Electron. Mater. **29**, 1067 (2000).

**Table 1.** Energetics of $(GaN)_x$ and $(GaN)_x Mn_2$ complexes. See Eqs. (1) - (3) for definitions.

| x | $E_b^{(1)}$ (eV) | $\Delta E$ (eV) | $\Delta E_0$ (eV) |
|---|---|---|---|
| 1 | 0 | 5.39 | - |
| 2 | 2.01 | 4.41 | 4.02 |
| 3 | 2.56 | 6.29 | 3.66 |

**Figure Captions**

Fig. 1. Ground state cluster geometries of $(GaN)_x$ (left panel) and $(GaN)_xMn_2$ (right panel). The magnetic moment of each cluster is also provided.

Fig. 2. Total density of states of $(Ga_{14}Mn_2)N_{16}$ supercell for majority spin (top) and minority spin (bottom).

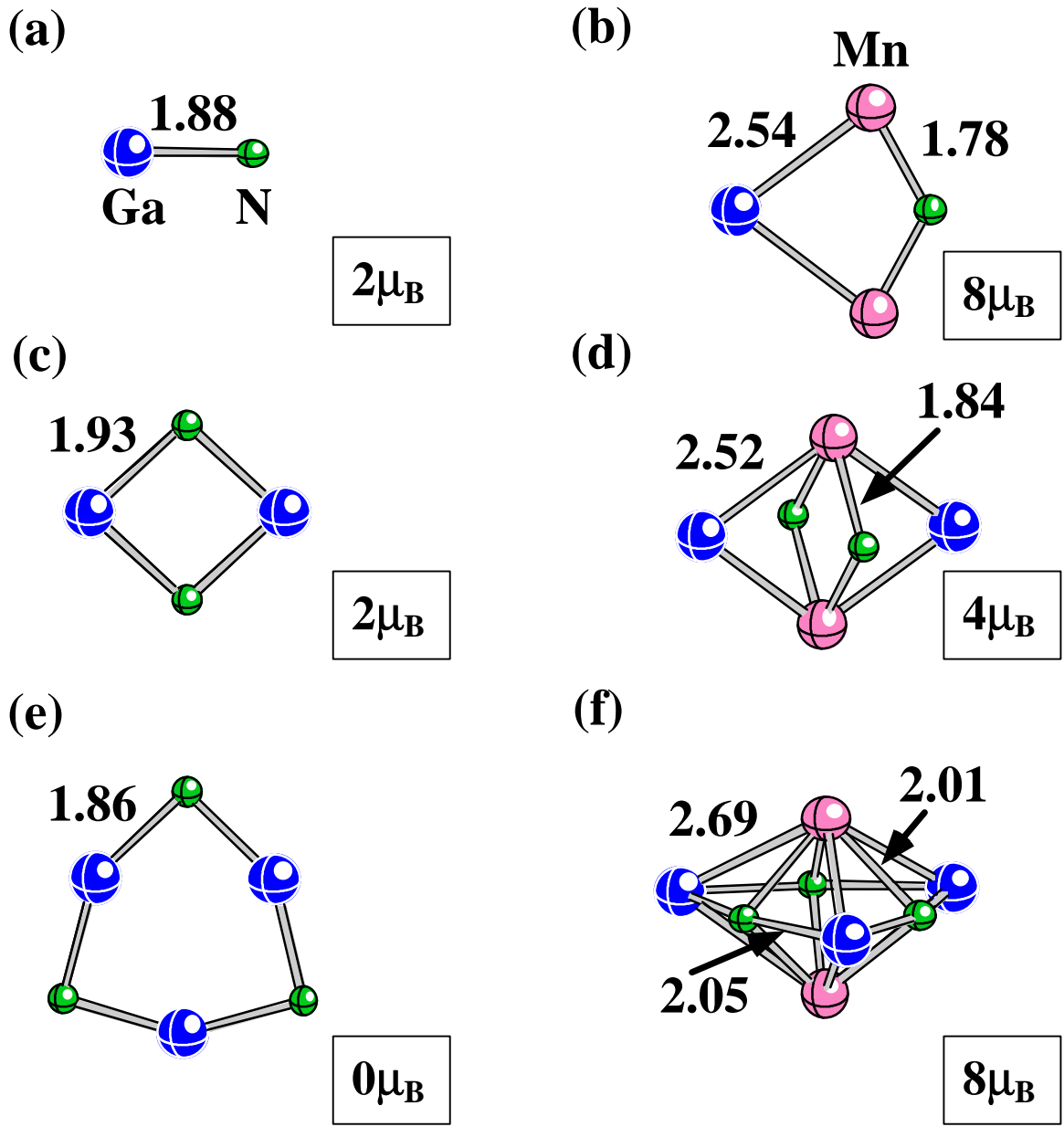

Fig. 1

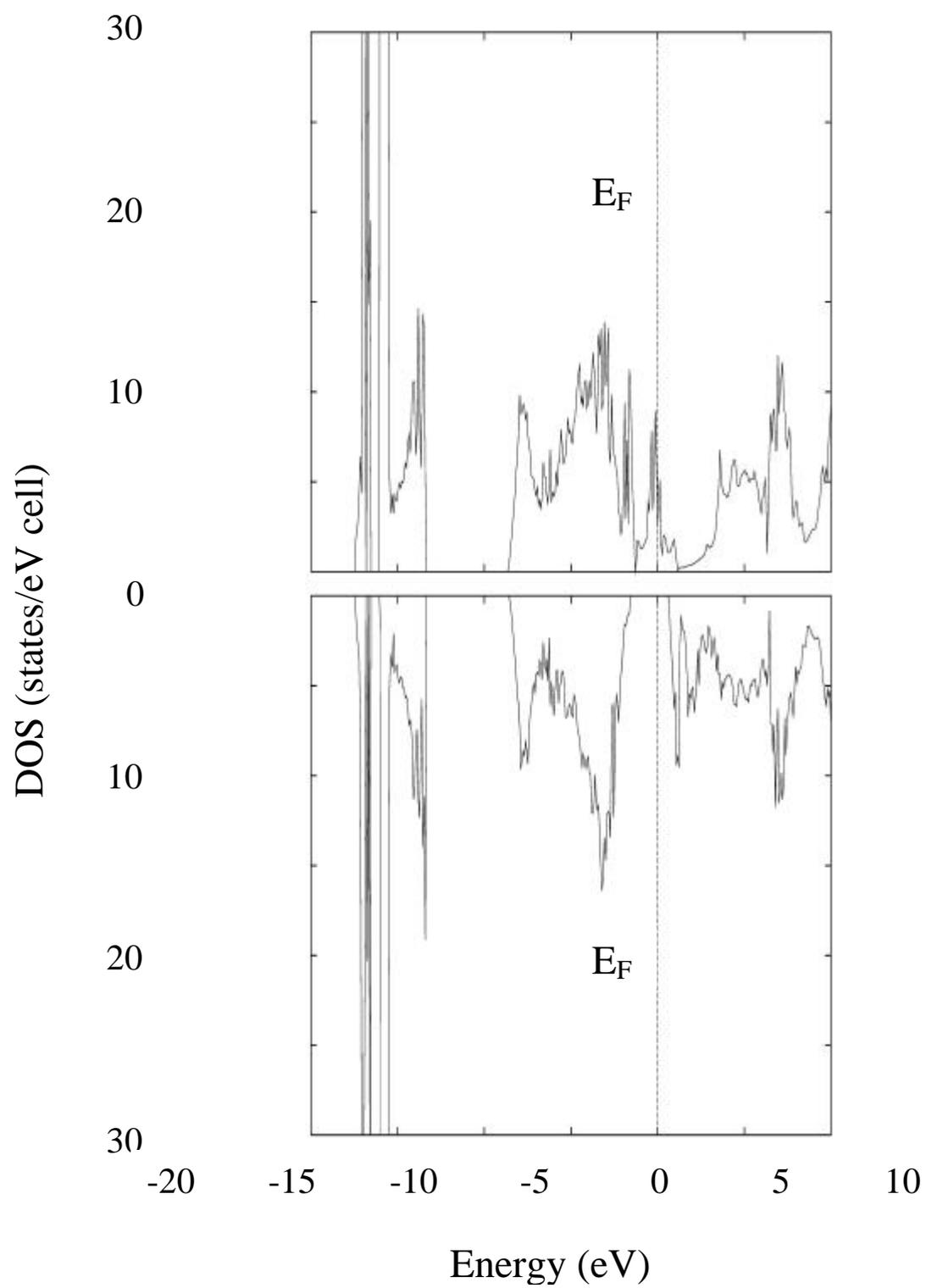

Fig. 2